\documentclass[manuscript]{aastex}

\slugcomment{}

\shorttitle{Flux rope stability}
\shortauthors{Joshi et al.}

\begin{document}


\title{Formation of Compound Flux Rope by The Merging of Two Filament Channels, Associated Dynamics and its Stability}


\author{Navin Chandra Joshi\altaffilmark{1}}
	\affil{School of Space Research, Kyung Hee University, Yongin, Gyeonggi-Do, 446-701, Korea; navin@khu.ac.kr, njoshi98@gmail.com}
\author{Tetsuya Magara\altaffilmark{2}}
	\affil{Department of Astronomy and Space Science, School of Space Research, Kyung Hee University, Yongin, Gyeonggi-Do, 446-701, Korea}
\author{Satoshi Inoue\altaffilmark{1}}
	\affil{School of Space Research, Kyung Hee University, Yongin, Gyeonggi-Do, 446-701, Korea}


\altaffiltext{1}{School of Space Research, Kyung Hee University, Yongin, Gyeonggi-Do, 446-701, Korea}
\altaffiltext{2}{Department of Astronomy and Space Science, School of Space Research, Kyung Hee University, Yongin, Gyeonggi-Do, 446-701, Korea}


\begin{abstract}

We present the observations of compound flux rope formation via merging of two nearby filament channels, associated dynamics and its stability that occurred on 2014 January 1 using multiwavelength data. We have also discussed the dynamics of cool and hot plasma moving along the newly formed compound flux rope. The merging started after the interaction between the southern leg of northward filament and the northern leg of the southward filament at $\approx$01:21 UT and continue until a compound flux rope formed at $\approx$01:33 UT. During the merging the cool filaments plasma heated up and started to move along the both side of the compound flux rope i.e., toward north ($\approx$265 $km~s^{-1}$) and south ($\approx$118 $km~s^{-1}$) from the point of merging. After traveling a distance of $\approx$150 Mm towards north the plasma become cool and started to returns back towards south ($\approx$14 $km~s^{-1}$) after $\approx$02:00 UT. The observations provide an clear example of compound flux rope formation via merging of two different flux ropes and occurrence of flare through tether cutting reconnection. However, the compound flux rope remained stable in the corona and made an confined eruption. The coronal magnetic field decay index measurements revealed that both the filaments and the compound flux rope axis lies within the stability domain (decay index $< 1.5$), which may be the possible cause for their stability. The present study also deals with the relationship between the filaments chirality {(sinistral)} and the helicity (positive) of the surrounding flux rope.

\end{abstract}




\section{Introduction}
\label{}
Solar eruptive events are magnetic transient events on the Sun, that occur due to release of magnetic energy via magnetic reconnection. These are solar flares, prominences/filament eruptions, jets, surges, coronal mass ejections (CMEs), etc. (see review articles by Benz 2008; Shibata \& Magara 2011, references cited therein). Filaments are the cool and dark plasma material suspended in the hot low laying corona with the support of magnetic field (Mackay et al. 2010; Labrosse et al. 2010). Filaments are known to lie in the dips of the surrounded magnetic flux ropes. These magnetic flux ropes become visible in the corona when the hot plasma moves along it (Li \& Zhang 2013; Joshi et al. 2014). Filament can have either sinistral or dextral chirality, depending on the axial magnetic field directions and can be estimated using various observations (Martin 1998). Association of filament chirality and helicity of the surrounding magnetic field has been reported in the last few decades (Rust \& Kumar 1994; Martin 1998; Chae 2000; Pevtsov et al. 2003; Martin 2003;
Muglach et al. 2009; Chandra et al. 2010; Joshi et al. 2014). Two nearby filaments sometimes show the merging dynamics when they come closer to each other due to magnetic shearing. Some authors reported observations showing this kind of merging (Schmieder et al. 2004), while some other produce numerical simulations and presented various condition of filament merging (DeVore et al. 2005; Aulanier et al. 2006).

Filaments are generally formed due to the reconfiguration of coronal field lines due to flux emergence, cancellation, various kinds of surface motions and due to the emergence of toroidal magnetic field (Rust 2001; Magara \& Longcope 2003). These filaments undergo eruptions due to the imbalance between upward magnetic pressure and downward magnetic tension and gravity forces. After the eruptions, these structures give huge CMEs, which are the ejection of plasma and magnetic field from the Sun into interplanetary space (Chen 2011; Forbes 2000). These CMEs are mainly responsible for the interplanetary consequences as well as near Earth disturbances (Joshi et al. 2013b). Several efforts have been carried out to understand the physical mechanism behind the eruption of filaments as well as the flux ropes (Chen 2011, references cited therein). Shearing and/or twisting of magnetic field lines are known to be responsible for the eruption (Hagyard et al. 1984; Fan \& Gibson 2004; Inoue et al. 2011; Sun et al. 2012). Magnetic flux emergence and cancellation has also been reported as a cause for these eruptions (Magara \& Longcope 2003; Magara 2006;Archontis \& T{\"o}r{\"o}k 2008). Ideal magnetohydrodynamics instabilities such as "kink" and "torus" instabilities are also known to plays an key role for the global eruption of theses structures (T{\"o}r{\"o}k et al. 2004; T{\"o}r{\"o}k \& Kliem 2005; Kliem \& T{\"o}r{\"o}k 2006; Srivastava et al. 2010, 2013). Several models such as "tether cutting","magnetic breakout" and various numerical simulations have also been given in order to explain the eruption phenomena (Antiochos et al. 1999; Moore et al. 2001; DeVore \& Antiochos 2008; D{\'e}moulin \& Aulanier 2010; Fan 2010; Chen 2011; Inoue et al. 2014).

Sometime even after the strong disturbances, the filaments as well as the flux ropes show confined eruptions that later settle down to its original positions. These are called failed or confined eruptions and have been studied (Ji et al. 2003; T{\"o}r{\"o}k \& Kliem 2005; Liu et al. 2009; Kumar et al. 2011; Shen et al. 2012; Kuridze et al. 2013; Joshi et al. 2013a; Filippov 2013; Joshi et al. 2014). The overlying coronal magnetic field know to play and important role for these kind of stable eruptions (Ji et al. 2003; T{\"o}r{\"o}k \& Kliem 2005; Liu et al. 2009; Guo et al. 2010; Kumar et al. 2011; Kuridze et al. 2013; Joshi et al. 2014). Ji et al. (2003), studied an failed eruption and, interpreted that the overlying closed magnetic field was responsible for the failed ejection. T{\"o}r{\"o}k \& Kliem (2005), modeled the confined as well as eruptive kink-unstable flux ropes and interpreted that the decrease of the overlying coronal magnetic field with height is an key factor that decides the confinement or eruption of the flux ropes. Liu et al. (2009), presented observations of failed filament eruption, and found that the asymmetric coronal background field can also influence the full eruption of filaments. Guo et al. (2010), studied the onset and stability conditions of a confined eruption, and found that the value of decay index was below the critical value for the torus instability to occur at above the erupting flux rope. Kuridze et al. (2013), also studied the failed filament eruption and found that the overlying large scale, magnetic loop arcade may caused the confinement of this eruption. Recently, Joshi et al. (2014) reported observations of an confined eruption of filament and associated flux rope and found that the whole filament axis lies within the region where the values of decay index was less than one. This caused the confined ejection of the filament and flux rope. Apart from these observational studies the numerical simulations have also been carried out in order to understand the stability (T{\"o}r{\"o}k \& Kliem 2005; Guo et al. 2010; D{\'e}moulin \& Aulanier 2010; An \& Magara 2013). Filaments sometime also show the reformation dynamics mainly after the confined or partial failed eruptions (Gilbert et al. 2007; Koleva et al. 2012; Joshi et al. 2014).

In the present work, we have presented the multiwavelength observations of compound flux rope formation, associated plasma dynamics and its stability on 2014 January 1. We have also made an attempt to find the relationship between the filament chirality and surrounding flux rope helicity. Observational data set used in the present work are described in the Section 2. Morphological interpretations of the event in different wavelengths are presented in the Section 3. Relationship between the filament chirality and helicity of surrounding magnetic structure is presented in the Section 4. Section 5 deals with the description of the coronal magnetic field measurements. Results and their discussions are presented in the final section.


\section{Observational Data Set}
\label{}
$H\alpha$ observations for the current study are collected from {\it Global Oscillation Network Group} (GONG) network archive (i.e., http://halpha.nso.edu/archive.html). GONG observe $H\alpha$ images at seven sites over the globe in 6563 \AA~wavelength with spatial resolution of 1$"$ and cadence of around 1 min (Harvey et al. 2011). The $H\alpha$ observations are used to study the dynamics of cool filament plasma along the magnetic flux rope. {\it Atmospheric Imaging Assembly} (AIA) and {\it Helioseismic and Magnetic Imager} (HMI;  Schou et al.(2012)) are two instruments on-board {\it Solar Dynamics Observatory} (SDO; Pesnell et al.(2012)). AIA observes the full solar disk in three Ultra-Violet (UV) continuum wavelengths and seven Extreme Ultra-violet (EUV) narrow band wavelengths (Lemen et al. 2012). SDO/AIA has a pixel size of 0.6$"$ and minimum cadence of 12 seconds. Therefore, it provides multi-temperature and high-resolution observations of the whole Sun. HMI provides full disk magnetogram images of the Sun at 1$"$ resolution with a minimum cadence of 45 seconds. We have used SDO/HMI magnetogram data to investigate the magnetic configuration around the filaments and flux rope as well as to estimate the values of decay index at different heights in the solar corona over the flux rope. To see the X-ray sources during the flare, we reconstructed X-ray images from the {\it Reuven Ramaty High-Energy Solar Spectroscopic Imager} (RHESSI; Lin et al. (2002)). We reconstructed the X-ray images in the two energy bands viz., 6-12 and 12-25 keV, using collimators from 3F to 9F and the CLEAN algorithm (Hurford et al. 2002).


\section{Multiwavelength Observations of Merging of Filament Channels and Associated Plasma Dynamics}
\label{}

Figure 1 shows the full disk images from GONG in $H\alpha$ wavelength (left panel) as well as from SDO/AIA in 304 \AA~wavelength (right panel) on January 1, 2014 at 00:00 UT. Boxes in each figures show the locations of the filaments. It is evident from the Figure 1 that the filaments lie near the disk center (i.e., $\approx$S09 E09) and oriented along the north-south direction. There was an small active region NOAA AR 11938 situated near to the filaments. Figure 2 shows the closed view of the region under study in this work. The panels (a) and (b) show the existence of two different filaments and associated magnetic channels in AIA 304 \AA~and 171 \AA~wavelength images respectively. Panel (c) shows the SDO/HMI images overplotted by the tracked filaments visible in the AIA images. It is clear that there is a crossing between the southern leg of northward filament and the northern leg of southward filaments. The norther leg of the southward filament located under the southern leg of northward filaments, which makes it less visible in on disk projection. In the upcoming subsections we will discuss the merging of two filaments, and associated magnetic field channels, dynamical motion of hot and cool plasma along the newly flux rope in different wavelengths using GONG $H\alpha$ and SDO/AIA EUV observations.


\subsection{$H\alpha$, X-ray and EUV Observations} 
\label{}

Figure 3, represents the selected images showing the dynamics of cool filaments plasma in $H\alpha$ wavelength observed by GONG. $H\alpha$ images provide crucial information regarding the chromospheric region of the Sun as well as the filament dynamics. Figures 4,5 and 6 show the selected images displaying the dynamics of the filament plasma as well as the flux rope in 304, 171 and 131 \AA~EUV images respectively observed by SDO/AIA. SDO/AIA 304 \AA~images provides information at chromospheric and transition region, while the SDO/AIA 171 and 131 \AA~images give quiet corona region and flaring regions respectively. The initial brightening appears near the junction of two filaments legs at $\approx$01:21 UT (c.f., Figures 3a, 4a, 5a, 6a, 7a, 7e and 7i). This brightening may be due to the reconnection between the low laying field lines of two filament channels. Then after the field lines of the two filament channels get reconnected and merged continuously during $\approx$01:21 to $\approx$01:33 UT (c.f., Figure 7 and zoomed-aia-171-movie.mpeg, zoomed-aia-304-movie.mpeg and zoomed-aia-131-movie.mpeg). Figure 7 displays the selected SDO/AIA 171, 304 and 131 \AA~wavelengths images showing the merging of two filament channels field lines (i.e., flux ropes) and the formation of new flux rope. After this, the newly formed flux rope will be referred as "compound flux rope" through the paper. The brightening in the merging region also gets increases which may be due to the heating via continuous merging/reconnection of field lines. During this process we also observed an C class flare at the merging site, starts at $\approx$01:27 UT (c.f., Figures 7c, 7g and 7k). The flare peaked at $\approx$01:33 UT and ended at $\approx$01:47 UT (as per the NOAA record of SF class flare). Soon after the flare triggers we have observed the formation of two ribbons at around $\approx$01:31 UT in SDO/AIA 1600 \AA~wavelength (c.f., Figure 8 and zoomed-aia-1600-movie.mpeg). The sudden brightening and the formation of two ribbon is also observed in GONG $H\alpha$ images (c.f., Figures 3b-3c). It is observed from Figures 8a-8b that both the flare ribbons lie in the different polarity regions. Southern and northern ribbons lie on the negative and positive magnetic polarity respectively. In between the flare ribbons we have also observed the RHESSI 6-12 and 12-25 keV X-ray sources (c.f., Figure 3b, 3c and 8c). From these observations the triggering of flare may be well interpreted using tether cutting model via filament channels field lines reconnection. In Figure 9 we have plotted a cartoon of the reconnection scenario.

Simultaneously with the merging, the filament cool plasma gets heated up and started to move towards northward and southward directions along the compound flux rope, (c.f., Figures 3b, 4b, 5b and 6b). The heated plasma moved towards north and south with an average speeds of around $\approx$265 $km~s^{-1}$ and $\approx$118 $km~s^{-1}$ respectively. Figure 10a show the distance time plot of the northward (black curve) and southward (blue curve) moving hot plasma constructed using SDO/AIA 171 \AA~wavelength images. The red color shows the GOES soft X-ray curve at 1.0-8.0 \AA~wavelength. On comparing GOES curve with the distance-time profiles, it is observed that the hot filament material moved toward the north and south while the flare is taking place simultaneously. The approximate trajectories along which this measurements have been performed are shown by the dotted black and blue lines in Figure 5e. During the northward motion of hot plasma, the filament disappeared, which may be due to the heating and shift of $H\alpha$ center absorption line. The filament disappears during $\approx$01:30-01:45 UT in $H\alpha$ images (c.f., Figures 3b-3d). The hot plasma continue to move towards north along the northern part of the compound flux rope and observed clearly in EUV wavelengths during $\approx$01:24 to $\approx$01:35 UT (c.f., Figures 4b-4d, 5b-5e and 6b-6e). High resolution SDO/AIA EUV images provide opportunity to see the surrounding compound flux rope after the hot plasma fill the whole flux rope. The complete s-shaped sigmoid structure of the compound flux rope is visible at $\approx$01:39 UT (c.f., Figures 4e, 5f and 6f). Along with the northward motion the surrounding flux rope also gets twisted right handedly and can be seen in EUV channels (c.f., Figures 4e-4f, 5e-5f and 6e-6f). After reaching an distance of around $\approx$150 Mm towards north, some cool as well as hot material moves towards the northern foot point of the compound flux rope during $\approx$01:35 UT to $\approx$01:44 UT with an average speed of around $\approx$150 $km~s^{-1}$ (c.f., Figures 4f, 5g and 6g). Figure 10b shows the distance time plot of the moving plasma towards the northern footpoint of the compound flux rope, estimated using AIA 171 \AA~images. The approximate trajectory along which this measurement have been performed are shown by the dotted black line in Figure 5g. High resolution images of SDO/AIA 171 \AA~also shows its various northern and southern foot points (c.f., Figures 4f, 5g and 6g). It is also evident that the northern/southern footpoints of the compound flux rope lies in the positive/negative polarity regions (c.f., Figures 4e, 5f and 6f). The appearance of compound flux rope in all the EUV channels show that the flux rope consist multi-thermal plasma. Complex fine structure of the compound flux rope can be seen in the high resolution AIA images (c.f., Figures 4g-4h, 5g-5h and 6g-6h). 

The filaments as well as compound flux rope not erupted even after the strong disturbance during the filament channels merging and triggering of flare. After reaching a distance of $\approx$150 Mm on disk towards north, most of the filament material stops and started returning/restoring southward motion along the compound flux rope (c.f., Figures 3g-3h and 4g-4h). This returning motion started at $\approx$02:00 UT. The returning motion continues along the same magnetic field channel and reformed the filament on the northern part of the compound flux rope (c.f., Figures 3i and 4i) with an average slow speed of $\approx$14 $km~s^{-1}$. Figure 10c shows the distance time plot of the moving plasma towards the southward along the compound flux rope, estimated using GONG $H\alpha$ images. The approximate trajectory along which this measurements have been performed are shown by the dotted white line in Figure 3i. Along with the formation of northern filament, the southern filament also appeared in the southern part of the compound flux rope. Reformed filaments can be seen in the $H\alpha$ image at around 03:36:54 UT (c.f., Figure 3i). Although the reformation and reappearance of cool plasma (i.e., filaments) continue up to several minutes after $\approx$03:36 UT also. The overall dynamics can also be clearly seen in $H\alpha$ and SDO/AIA movies (i.e., h-alpha-movies.mpeg, aia304movie.mpeg, aia171movie.mpeg and aia131movie.mpeg).

From the observations seems that the compound flux rope moves upwards, remains stable and returns to the original position (see animations in Figures 4,5 and 6). Confined erupted compound flux rope also disturbed the overlying corona arcades, as a result the arcade started to oscillate with some small amplitudes (c.f., Figures 5c-g and AIA 171 movie). This also gives the signature to confined upward eruption of the stable compound flux rope. However, the exact height of the flux rope is hard to determine due to the on disk projection. It is also evident that as the hot plasma material became cooled down, the flux rope also become invisible simultaneously (c.f., Figures 5i and 6i). For more detail please refer to the SDO/AIA 171 and 131 \AA~movies (i.e., aia171movie.mpeg and aia131movie.mpeg). In order to help to understand the plasma dynamics of the event, we have made a cartoon shown in Figure 11.


\section{Chirality and Helicity of Filament and Flux Rope}
\label{}

In this section, we will discuss the observational manifestations of filaments chirality and helicity of surrounding compound flux rope. Figure 12(a) shows one selected GONG $H\alpha$ images few hour before the event i.e., at 22:42:34 UT on 2013 December 31, in order to study the chirality of the filaments. It is clear from this image that the fine threads and fibrils are oriented leftward from the filaments axis, that provides evidence that the chirality of the filaments should be sinistral (Martin 1998). Figure 12(b) represents the SDO/AIA 171 \AA~image overplotted by SDO/HMI magnetogram at 01:00:11 UT. It shows that the southern footpoints of both the filaments lie on the negative polarity while the northern foot points lie on the positive polarity regions, which shows the evidence of sinistral chirality of filaments (Martin 1998). Also, the right-skewed alignment of overlying coronal arcade provides an key observational evidence showing the existence of the sinistral chirality (Martin 1998). From these observational facts we can confirm that the filaments has sinistral chirality.

Figure 13(a) represents the SDO/AIA 171 \AA~image at 01:28:11 UT showing that the bright hot plasma is moving behind the dark cool plasma material. In this way, we have observed a crossing of bright (below) and dark (above) material in EUV images. According to the Pevtsov et al. (2003) interpretation, our situation is of type I which corresponds to the positive helicity of flux rope. SDO/AIA 171, 131 and 94 \AA~wavelengths images clearly showing the right handed twist in the magnetic field of the compound flux rope, which shows the signature of positive helicity (c.f., Figures 13b-13d). Careful inspection shows that the flux rope has $\approx$2 turns around its axis (c.f., Figures 13b-13d). Also from the well-known hemispheric helicity rule the northern/southern hemispheres has negative/positive helicity (Pevtsov et al. 2003). Our observation is also in agreement with this rule. On the bases of these discussions we can say that, these observation confirms the relationship of sinistral chirality and positive helicity.


\section{Coronal Magnetic Field Measurements}
\label{}

To check the overlaying magnetic field conditions we have estimated the decay index over the compound flux rope axis. We have used the method of potential field extrapolation to estimate the decay index over the filament axis (Sakurai 1982). The decay index is defined as 
\begin{equation}
n=-\frac{\partial \log B}{\partial \log h},
\end{equation}
where B denotes the strength of the magnetic field and h is the height above the photosphere.
 
Figure 14 top panel shows the radial component of magnetic field (a) and the potential field extrapolated over it (b). Blue lines show the magnetic field lines. Figure 14 bottom panel shows the decay index measured over the compound flux rope and filaments axis at 86.4 Mm (c), 108 Mm (d) and 129.6 Mm (e). It is evident that the most of the filaments and compound flux rope axis lie in the region of decay index less then 1.5 at 86.4 Mm. Even at 108 Mm contour level 1.5 is partially appears near the flux rope axis. This results also suggest that the flux rope is stable against the torus instability if it located less than 86.4 Mm. Although, compound flux rope observed on the disk, so it is difficult to measure the on disk height. But, we used the overlying coronal arcade to estimate the approximate height of the flux rope. We have selected one potential field line matching best with the overlaying arcade and calculated the height of this potential field line from the photosphere (c.f. Figure 15(a)). The top of the potential field line comes out to be $\approx$86.4 Mm (c.f., Figure 15(b)). During the confined upward eruption, the compound flux rope was only able to disturbed the overlying arcade, as a result of this, the arcade started to oscillate with small amplitude. Hence, from this observation we can confirm that the flux rope does not reached up to a height of $\approx$86.4 Mm, which is the top height of the arcade (c.f., Figure 15). If we consider this height as the maximum height achieved by the compound flux rope during its confined eruption, than its axis will be at a height of $\approx$43.2 Mm. Although the actual top height of compound flux rope will be less then 86.4 Mm. From the decay index measurement it is evident that the flux rope remain stable below a height of around 86.4 Mm (c.f., Figure 14 bottom left panel), which is the assumed top height of the flux rope and the top height of coronal arcade (c.f., Figure 15). So the filament flux rope remain stable because it situated in the region of stability i.e., decay index $< 1.5$.


\section{Results and Discussions}
\label{}

We present high-resolution multiwavelength observations of the merging of filament channels, associated plasma dynamics and the stability of compound flux rope on 2014 January 1. The main results of the present study are as follows.

1. The observations clearly show the merging of two filament channels, triggering of solar flare and the formation of compound flux rope via tether cutting magnetic reconnection scenario in a nice manner.\\
2. The compound flux rope shows an stability in the corona because it does not reached at a height where the decay index is greater then 1.5, which makes it stable against the torus instability.\\
3. We have also found a clear association between the filaments chirality (sinistral) and magnetic helicity (positive) of surrounding flux rope.\\

The overall scenario of the flare and the formation of compound flux rope can be interpreted using tether cutting reconnection model (Moore et al. 2001). According to which, the initial onset of an explosion unleashed by internal tether-cutting reconnection between the sheared low laying field lines. Then after a new flux rope formed which either remain confined or eruptive (for detail please see Figure 1 of Moore et al.(2001) for reference). Based on the detailed analysis of the present event using multiwavelength observations, we have drawn schematics in order to explain the whole event (c.f., Figures 9 and 11). In our case, we found the signature of tether cutting reconnection in the form of brightening at the junction of two filament legs, that can be consider as the reconnection between the low laying sheared field lines (c.f., Figures 4a, 5a, 6a, 7, 9b and 11). The newly formed flux rope (i.e., compound flux rope) formed after this reconnection with the northward and southward movement of hot plasma along it (c.f., Figure 7 and associated animation), which then moved upward in the corona with the reconnection beneath of it (c.f., Figure 9b-c). This reconnection produce a solar flare with flare ribbons (c.f., Figure 8 and associated animation). The RHESSI 6-12 and 12-25 keV sourced formed near the merging area, which provides an strong signature of reconnection (c.f., Figures 7, 8c). The formation of flare loops (c.f., Figure 5g) at the point of merging provides the signature of formation of compact low laying field lines are in good agreement with the model (c.f., Figures 5g and 9c). Overall, the observations clearly show the tether cutting due to filament legs and associated filament channel field lines interaction, which allow the partial eruption of newly formed flux rope, triggering of flare underneath of it (please see schematics in Figures 9 and 11). However, the compound flux rope shows a confined eruption in the corona. To the best of our knowledge for the first time we have found a clear observation showing the filaments interaction/merging, formation of twisted compound flux rope, triggering of flare in one event. However, there are few other papers that deal with the observational signature of tether cutting reconnection via flux cancellation, between two low laying sheared arcades (Liu et al. 2010, 2013).

It has been reported in observations as well as in numerical simulations that the merging of two filaments are possible if it consist same axial magnetic fields (Schmieder
et al. 2004; DeVore et al. 2005; Aulanier et al. 2006). We have found a clear observations showing the merging of two filament having sinistral chirality (c.f., Figure 7 and zoomed-aia-171-movie.mpeg, zoomed-aia-304-movie.mpeg and zoomed-aia-131-movie.mpeg).

The initial onset conditions for these confined eruptions are also important to investigate. Liu et al. (2009), explained the initial onset was due to kink instability at unstable heights, while Kuridze et al. (2013), explained the magnetic breakout scenario as the onset condition for confined eruption. Recently, Joshi et al. (2014) explained that the eruption of eastern eruptive part destabilize the overlying magnetic field of the stable filament, which may allow the confined initial onset of it. However, in the present case for the first time we found that the tether cutting reconnection between the two filament legs was responsible for the initial onset of confined eruption of the compound flux rope (c.f., Figures  7, 8 and 9).

Fine structures of huge flux rope are rarely observed in solar corona. This is possible only after the hot plasma tracked the fine flux tubes and make them visible (Li \& Zhang
2013; Joshi et al. 2014). Recently, Li \& Zhang (2013), examine the fine structure of two flux ropes and found that they are composed of $85\pm12$ and $102\pm15$ fine-scale structures having each ends with multiple footpoints. Our observation also show that the compound flux rope has several fine structures with both end of the flux rope consists of multiple foot points (c.f., Figures 4f, 5g-h, 6g-h and 13b-d).

The compound flux rope shows the stability against the eruption in the corona and make the eruption confined (c.f., Figure 3, 4, 5 and 6 and animations). For the stable equilibrium the value of decay index of the ambient magnetic field should not exceed the critical value i.e., 1 for straight current channel and 1.5 for circular current channel depending on the different flux rope geometry (Filippov \& Den 2001; D{\'e}moulin \& Aulanier 2010). To check the stability we have calculated the decay index near the compound flux rope axis and found that the whole axis lies in the region where the value of decay index is less than the critical value (i.e., less than 1.5) up to 86.4 Mm. Our measurements also suggest that the maximum height of the flux rope as well as its axis should be less then 86.4 Mm and 43.2 Mm coronal heights respectively (see Section 5 for detail). Hence, the overlying coronal magnetic field playing an crucial role for the stability of filaments and compound flux rope against torus instability. Also, for the kink instability to erupt an flux rope should contain 2-3 turns (T{\"o}r{\"o}k et al. 2004). But in our observations, we have found that the flux rope has less then two turns (c.f., Figure 13 b-d), which make the flux rope stable against the kink instability also. In future, In order to understand more clearly the flux rope merging, reconnection and the stability, we are going to simulate these kind of stable flux rope eruptions in the solar corona. 

It is observed that in general the northern/southern filaments have dextral/sinistral chirality and hence the negative/positive helicity respectively (Chae 2000; Pevtsov et al.
2003). However some authors also reported the signature of reverse characteristics as well as the evidence of mixed helicity in the solar corona (Pevtsov et al. 2003; Martin 2003;
Muglach et al. 2009). Recently, Joshi et al. (2014), also reported the relationship between the filament chirality (sinistral) and helicity (positive) of flux rope and also found an good correlation. In the present work, we have also discussed the relationship between filament chirality and associated flux rope helicity and found a clear relationship. Using various observational manifestations we obtain that the filaments has sinistral chirality (c.f., Section 4 and Figure 12 for more detail). From the apparent motion of plasma within the compound flux rope as well as from the complete twisted geometry of flux rope in EUV images, it is confirm that the surround magnetic flux rope has positive helicity (c.f., Section 4 and Figure 13). Overall this observation is an another clear example showing the relationship between chirality and helicity.

More observational and simulation work are needed in order to understand the filaments interaction/merging dynamics in detail. The eruption of such big disk center filaments and flux ropes are mainly responsible for the fast Earth-directed CMEs and other interplanetary and near Earth consequences. However, sometime they remain stable under the coronal magnetic field conditions. Therefore, observational studies as well as numerical simulations of filament stabilization/eruption conditions may play an significant role to forecast the initial eruption conditions as well as space weather.


\acknowledgments
The authors thank referee for providing his constructive comments and suggestions. We thank SDO/AIA, SDO/HMI, NSO-GONG, GOES and RHESSI teams for providing their data for the present study. This work is supported by the BK21 plus program through the National Research Foundation (NRF) funded by the Ministry of Education of Korea. NCJ thank School of Space Research, Kyung Hee University for providing Postdoctoral grant. We also acknowledge the Basic Science Research Program (NRF-2013R1A1A2058705, PI: T. Magara) through the National Research Foundation of Korea (NRF) provided by the Ministry of Education, Science and Technology. SI was supported by the International Scholarship of Kyung Hee University.


\begin{thebibliography}{}

\bibitem[An \& Magara(2013)]{2013ApJ...773...21A} An, J.~M., \& Magara, T.\ 2013, \apj, 773, 21 
\bibitem[Antiochos et al.(1999)]{1999ApJ...510..485A} Antiochos, S.~K., DeVore, C.~R., \& Klimchuk, J.~A.\ 1999, \apj, 510, 485 
\bibitem[Archontis \& T{\"o}r{\"o}k(2008)]{2008A&A...492L..35A} Archontis, V., T{\"o}r{\"o}k, T.\ 2008, \aap, 492, L35 
\bibitem[Aulanier et al.(2006)]{2006ApJ...646.1349A} Aulanier, G., DeVore, C.~R., \& Antiochos, S.~K.\ 2006, \apj, 646, 1349 
\bibitem[Benz(2008)]{2008LRSP....5....1B} Benz, A.~O.\ 2008, Living Reviews in Solar Physics, 5, 1 
\bibitem[Chae(2000)]{2000ApJ...540L.115C} Chae, J.\ 2000, \apjl, 540, L115 
\bibitem[Chandra et al.(2010)]{2010SoPh..261..127C} Chandra, R., Pariat, E., Schmieder, B., Mandrini, C.~H., \& Uddin, W.\ 2010, \solphys, 261, 127 
\bibitem[Chen(2011)]{2011LRSP....8....1C} Chen, P.~F.\ 2011, Living Reviews in Solar Physics, 8, 1 
\bibitem[D{\'e}moulin \& Aulanier(2010)]{2010ApJ...718.1388D} D{\'e}moulin, P., \& Aulanier, G.\ 2010, \apj, 718, 1388 
\bibitem[DeVore \& Antiochos(2008)]{2008ApJ...680..740D} DeVore, C.~R., \& Antiochos, S.~K.\ 2008, \apj, 680, 740 
\bibitem[DeVore et al.(2005)]{2005ApJ...629.1122D} DeVore, C.~R., Antiochos, S.~K., \& Aulanier, G.\ 2005, \apj, 629, 1122 
\bibitem[Fan(2010)]{2010ApJ...719..728F} Fan, Y.\ 2010, \apj, 719, 728 
\bibitem[Fan \& Gibson(2004)]{2004ApJ...609.1123F} Fan, Y., \& Gibson, S.~E.\ 2004, \apj, 609, 1123 
\bibitem[Filippov(2013)]{2013ApJ...773...10F} Filippov, B.\ 2013, \apj, 773, 10
\bibitem[Filippov \& Den(2001)]{2001JGR...10625177F} Filippov, B.~P., \& Den, O.~G.\ 2001, \jgr, 106, 25177 
\bibitem[Forbes(2000)]{2000JGR...10523153F} Forbes, T.~G.\ 2000, \jgr, 105, 23153 
\bibitem[Gilbert et al.(2007)]{2007SoPh..245..287G} Gilbert, H.~R., Alexander, D., \& Liu, R.\ 2007, \solphys, 245, 287 
\bibitem[Guo et al.(2010)]{2010ApJ...725L..38G} Guo, Y., Ding, M.~D., Schmieder, B., et al.\ 2010, \apjl, 725, L38 
\bibitem[Hagyard et al.(1984)]{1984SoPh...91..115H} Hagyard, M.~J., Teuber, D., West, E.~A., \& Smith, J.~B.\ 1984, \solphys, 91, 115 
\bibitem[Harvey et al.(2011)]{2011SPD....42.1745H} Harvey, J.~W., Bolding, J., Clark, R., et al.\ 2011, Bulletin of the American Astronomical Society, 1745 
\bibitem[Hurford et al.(2002)]{2002SoPh..210...61H} Hurford, G.~J., Schmahl, E.~J., Schwartz, R.~A., et al.\ 2002, \solphys, 210, 61 
\bibitem[Inoue et al.(2014)]{2014ApJ...788..182I} Inoue, S., Hayashi, K., Magara, T., Choe, G.~S., \& Park, Y.~D.\ 2014, \apj, 788, 182 
\bibitem[Inoue et al.(2011)]{2011ApJ...738..161I} Inoue, S., Kusano, K., Magara, T., Shiota, D., \& Yamamoto, T.~T.\ 2011, \apj, 738, 161 
\bibitem[Ji et al.(2003)]{2003ApJ...595L.135J} Ji, H., Wang, H., Schmahl, E.~J., Moon, Y.-J., \& Jiang, Y.\ 2003, \apjl, 595, L135 
\bibitem[Joshi et al.(2014)]{2014ApJ...787...11J} Joshi, N.~C., Srivastava, A.~K., Filippov, B., et al.\ 2014, \apj, 787, 11 
\bibitem[Joshi et al.(2013)]{2013ApJ...771...65J} Joshi, N.~C., Srivastava, A.~K., Filippov, B., et al.\ 2013, \apj, 771, 65 
\bibitem[Joshi et al.(2013)]{2013AdSpR..52....1J} Joshi, N.~C., Uddin, W., Srivastava, A.~K., et al.\ 2013, Advances in Space Research, 52, 1 
\bibitem[Kliem \& T{\"o}r{\"o}k(2006)]{2006PhRvL..96y5002K} Kliem, B., T{\"o}r{\"o}k, T.\ 2006, Physical Review Letters, 96, 255002 
\bibitem[Koleva et al.(2012)]{2012A&A...540A.127K} Koleva, K., Madjarska, M.~S., Duchlev, P., et al.\ 2012, \aap, 540, A127 
\bibitem[Kumar et al.(2011)]{2011SoPh..272..301K} Kumar, P., Srivastava, A.~K., Filippov, B., Erd{\'e}lyi, R., \& Uddin, W.\ 2011, \solphys, 272, 301 
\bibitem[Kuridze et al.(2013)]{2013A&A...552A..55K} Kuridze, D., Mathioudakis, M., Kowalski, A.~F., et al.\ 2013, \aap, 552, A55 
\bibitem[Labrosse et al.(2010)]{2010SSRv..151..243L} Labrosse, N., Heinzel, P., Vial, J.-C., et al.\ 2010, \ssr, 151, 243 
\bibitem[Lemen et al.(2012)]{2012SoPh..275...17L} Lemen, J.~R., Title, A.~M., Akin, D.~J., et al.\ 2012, \solphys, 275, 17 
\bibitem[Li \& Zhang(2013)]{2013ApJ...770L..25L} Li, T., \& Zhang, J.\ 2013, \apjl, 770, L25 
\bibitem[Lin et al.(2002)]{2002SoPh..210....3L} Lin, R.~P., Dennis, B.~R., Hurford, G.~J., et al.\ 2002, \solphys, 210, 3 
\bibitem[Liu et al.(2013)]{2013ApJ...778L..36L} Liu, C., Deng, N., Lee, J., et al.\ 2013, \apjl, 778, L36 
\bibitem[Liu et al.(2010)]{2010ApJ...725L..84L} Liu, R., Liu, C., Wang, S., Deng, N., \& Wang, H.\ 2010, \apjl, 725, L84 
\bibitem[Liu et al.(2009)]{2009ApJ...696L..70L} Liu, Y., Su, J., Xu, Z., et al.\ 2009, \apjl, 696, L70 
\bibitem[Mackay et al.(2010)]{2010SSRv..151..333M} Mackay, D.~H., Karpen, J.~T., Ballester, J.~L., Schmieder, B., \& Aulanier, G.\ 2010, \ssr, 151, 333 
\bibitem[Magara(2006)]{2006ApJ...653.1499M} Magara, T.\ 2006, \apj, 653, 1499 
\bibitem[Magara \& Longcope(2003)]{2003ApJ...586..630M} Magara, T., \& Longcope, D.~W.\ 2003, \apj, 586, 630 
\bibitem[Martin(1998)]{1998ASPC..150..419M} Martin, S.~F.\ 1998, IAU Colloq.~167: New Perspectives on Solar Prominences, 150, 419 
\bibitem[Martin(2003)]{2003AdSpR..32.1883M} Martin, S.~F.\ 2003, Advances in Space Research, 32, 1883 
\bibitem[Moore et al.(2001)]{2001ApJ...552..833M} Moore, R.~L., Sterling, A.~C., Hudson, H.~S., \& Lemen, J.~R.\ 2001, \apj, 552, 833 
\bibitem[Muglach et al.(2009)]{2009ApJ...703..976M} Muglach, K., Wang, Y.-M., \& Kliem, B.\ 2009, \apj, 703, 976 
\bibitem[Pesnell et al.(2012)]{2012SoPh..275....3P} Pesnell, W.~D., Thompson, B.~J., \& Chamberlin, P.~C.\ 2012, \solphys, 275, 3 
\bibitem[Pevtsov et al.(2003)]{2003ApJ...595..500P} Pevtsov, A.~A., Balasubramaniam, K.~S., \& Rogers, J.~W.\ 2003, \apj, 595, 500 
\bibitem[Rust(2001)]{2001JGR...10625075R} Rust, D.~M.\ 2001, \jgr, 106, 25075 
\bibitem[Rust \& Kumar(1994)]{1994SoPh..155...69R} Rust, D.~M., \& Kumar, A.\ 1994, \solphys, 155, 69 
\bibitem[Sakurai(1982)]{1982SoPh...76..301S} Sakurai, T.\ 1982, \solphys, 76, 301 
\bibitem[Schmieder et al.(2004)]{2004SoPh..223..119S} Schmieder, B., Mein, N., Deng, Y., et al.\ 2004, \solphys, 223, 119 
\bibitem[Schou et al.(2012)]{2012SoPh..275..229S} Schou, J., Scherrer, P.~H., Bush, R.~I., et al.\ 2012, \solphys, 275, 229 
\bibitem[Shen et al.(2012)]{2012ApJ...750...12S} Shen, Y., Liu, Y., \& Su, J.\ 2012, \apj, 750, 12 
\bibitem[Shibata \& Magara(2011)]{2011LRSP....8....6S} Shibata, K., \& Magara, T.\ 2011, Living Reviews in Solar Physics, 8, 6 
\bibitem[Srivastava et al.(2013)]{2013ApJ...765L..42S} Srivastava, A.~K., Erd{\'e}lyi, R., Tripathi, D., et al.\ 2013, \apjl, 765, L42 
\bibitem[Srivastava et al.(2010)]{2010ApJ...715..292S} Srivastava, A.~K., Zaqarashvili, T.~V., Kumar, P., \& Khodachenko, M.~L.\ 2010, \apj, 715, 292 
\bibitem[Sun et al.(2012)]{2012ApJ...748...77S} Sun, X., Hoeksema, J.~T., Liu, Y., et al.\ 2012, \apj, 748, 77 
\bibitem[T{\"o}r{\"o}k \& Kliem(2005)]{2005ApJ...630L..97T} T{\"o}r{\"o}k, T., \& Kliem, B.\ 2005, \apjl, 630, L97 
\bibitem[T{\"o}r{\"o}k et al.(2004)]{2004A&A...413L..27T} T{\"o}r{\"o}k, T., Kliem, B., \& Titov, V.~S.\ 2004, \aap, 413, L27 

\end{thebibliography}

\clearpage
\begin{figure}
\vspace*{-5cm}
\centerline{
	\hspace*{0.0\textwidth}
	\includegraphics[width=1.2\textwidth,clip=]{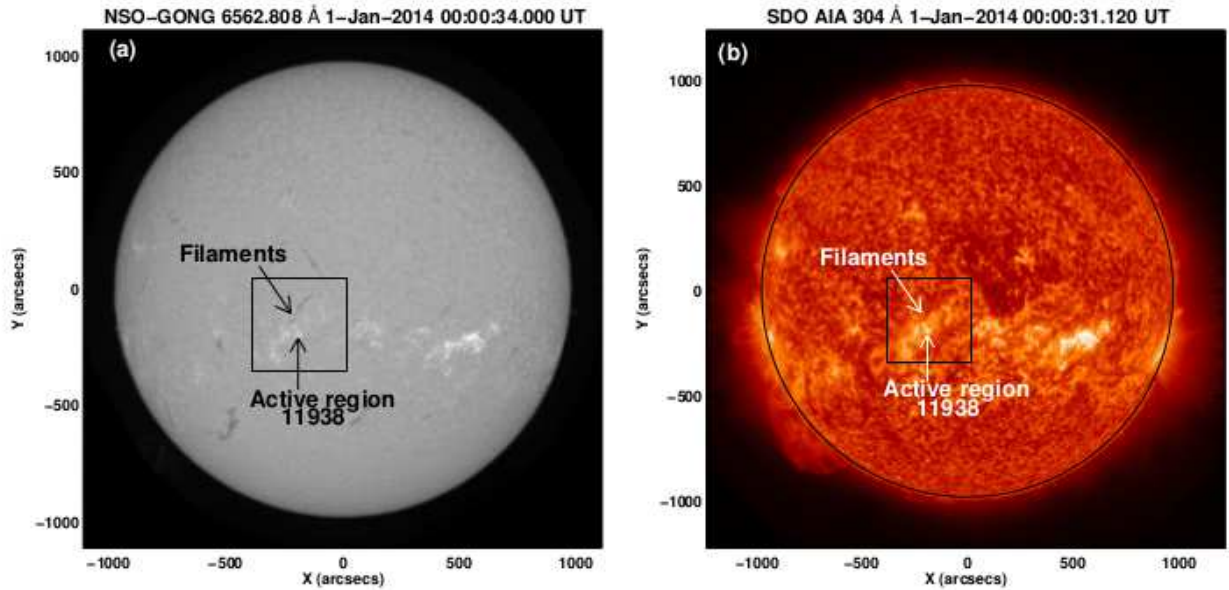}
	}
\vspace*{-9cm}
\caption{Full disk Global Oscillation Network Group (GONG) $H\alpha$ image (a) and full disk SDO/AIA 304 \AA~image (b) on 2014 January 1 at 00:00 UT. Boxes in each figures show the locations of the filaments under study. The nearby active region NOAA AR 11938 is also shown by the arrow.}
\label{}
\end{figure}
\clearpage
\begin{figure}
\vspace*{-5cm}
\centerline{
	\hspace*{0.0\textwidth}
	\includegraphics[width=1.2\textwidth,clip=]{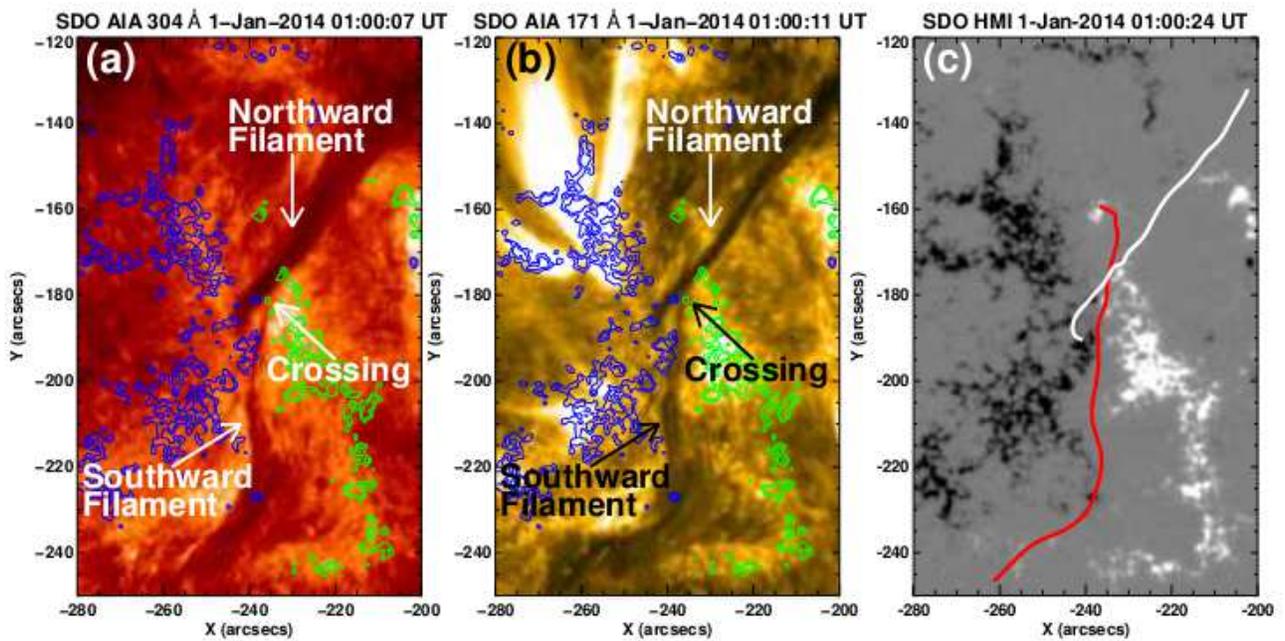}
	}
\vspace*{-9cm}
\caption{Zoomed SDO/AIA 304 (a) and 171 \AA~(b) images overplotted with SDO/HMI contours at 01:00 UT, showing the crossing of northward and southward filaments legs. The green/blue contours show the positive/negative polarity regions. The contour levels are $\pm200$, $\pm400$ Gauss. (c) SDO/HMI magnetogram overplotted by the filaments axis tracked using SDO/AIA 304 and 171 \AA~ images shown in panels (a) and (b).}
\label{}
\end{figure}
\clearpage
\begin{figure}
\vspace*{-2cm}
\centerline{
	\hspace*{0.0\textwidth}
	\includegraphics[width=1.2\textwidth,clip=]{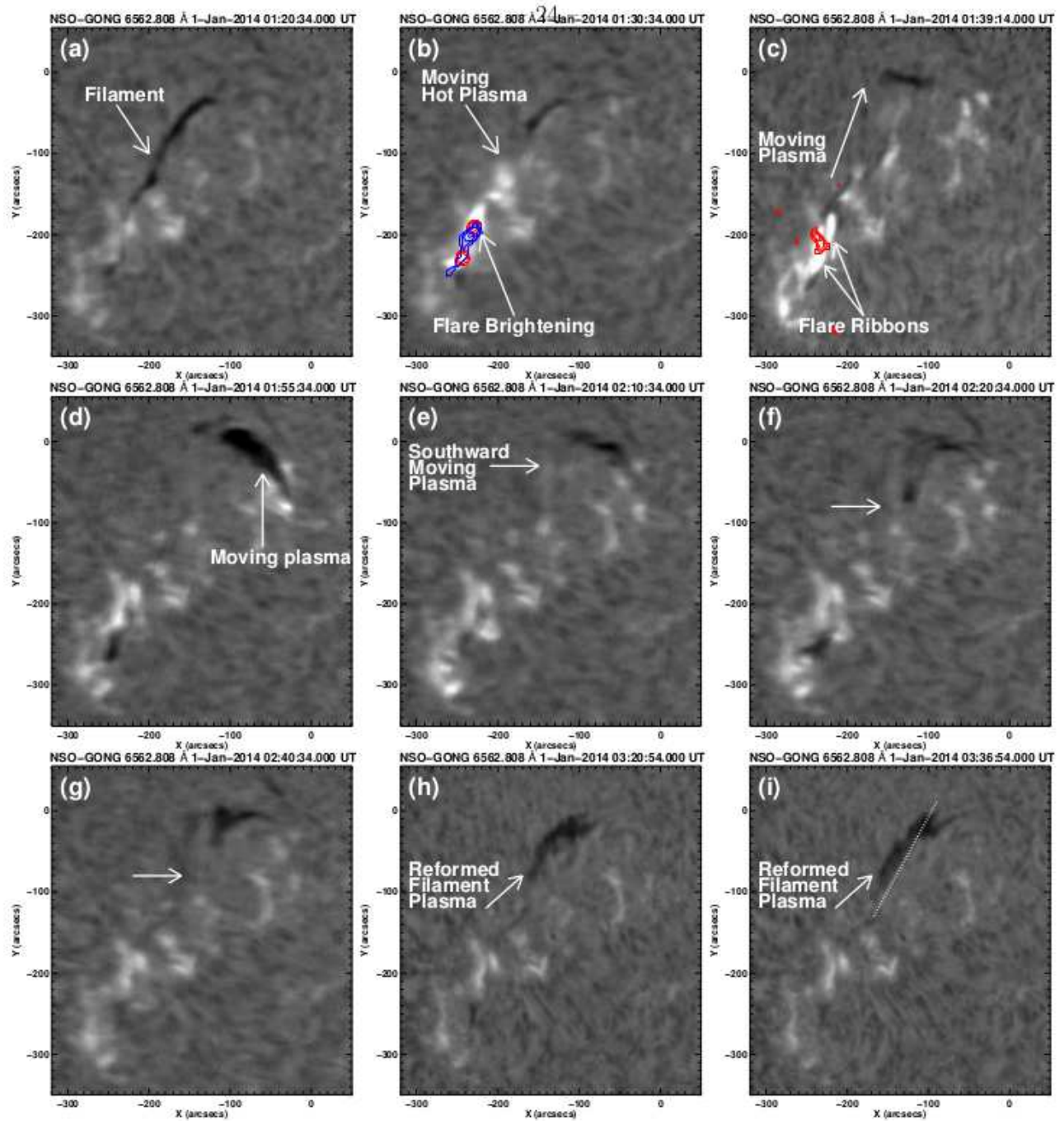}
	}
\vspace*{-4cm}
\caption{Selected GONG H$\alpha$ images showing the dynamics of the cooler filament plasma. Dotted white line in panel (i) represent the approximate trajectory along which the 
distance-time plot of southward moving cool plasma has been measured. The red and blue contours are the RHESSI X-ray contours at 6-12 and 12-25 keV energy bands. The contour levels are 60\%,70\%,80\%, 95\% of peak intensity. The integration time is 20 seconds.} 
\label{}
\end{figure}
\clearpage
\begin{figure}
\vspace*{-2cm}
\centerline{
	\hspace*{0.0\textwidth}
	\includegraphics[width=1.2\textwidth,clip=]{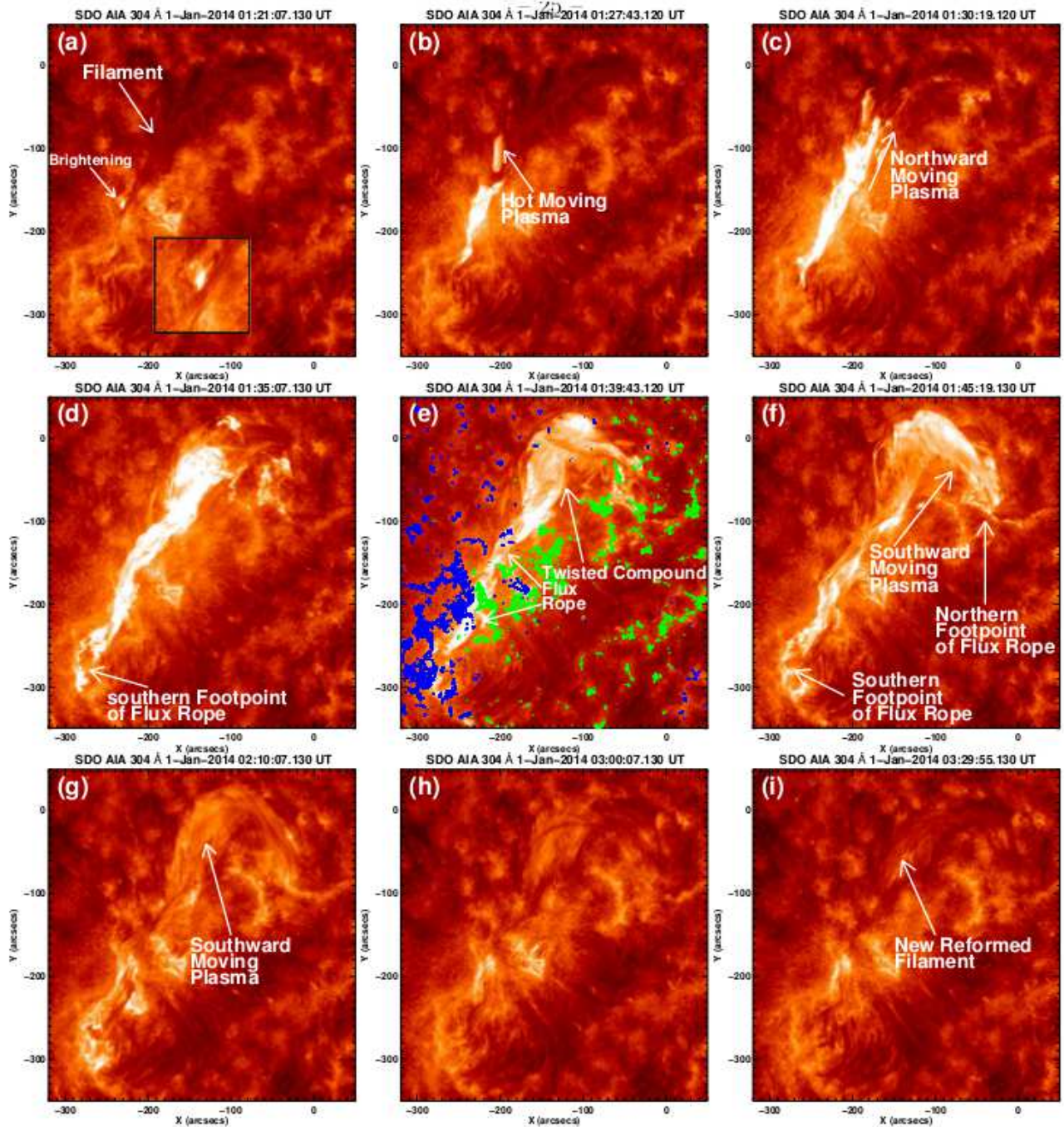}
	}
\vspace*{-4cm}
\caption{Selected SDO/AIA 304 \AA~images showing the overall dynamics of plasma at chromospheric temperature. The green/blue contours in panel (e) shows the positive/negative polarity regions. The contour levels are $\pm100$, $\pm200$, $\pm400$ Gauss.}
\label{}
\end{figure}
\clearpage
\begin{figure}
\vspace*{-2cm}
\centerline{
	\hspace*{0.0\textwidth}
	\includegraphics[width=1.2\textwidth,clip=]{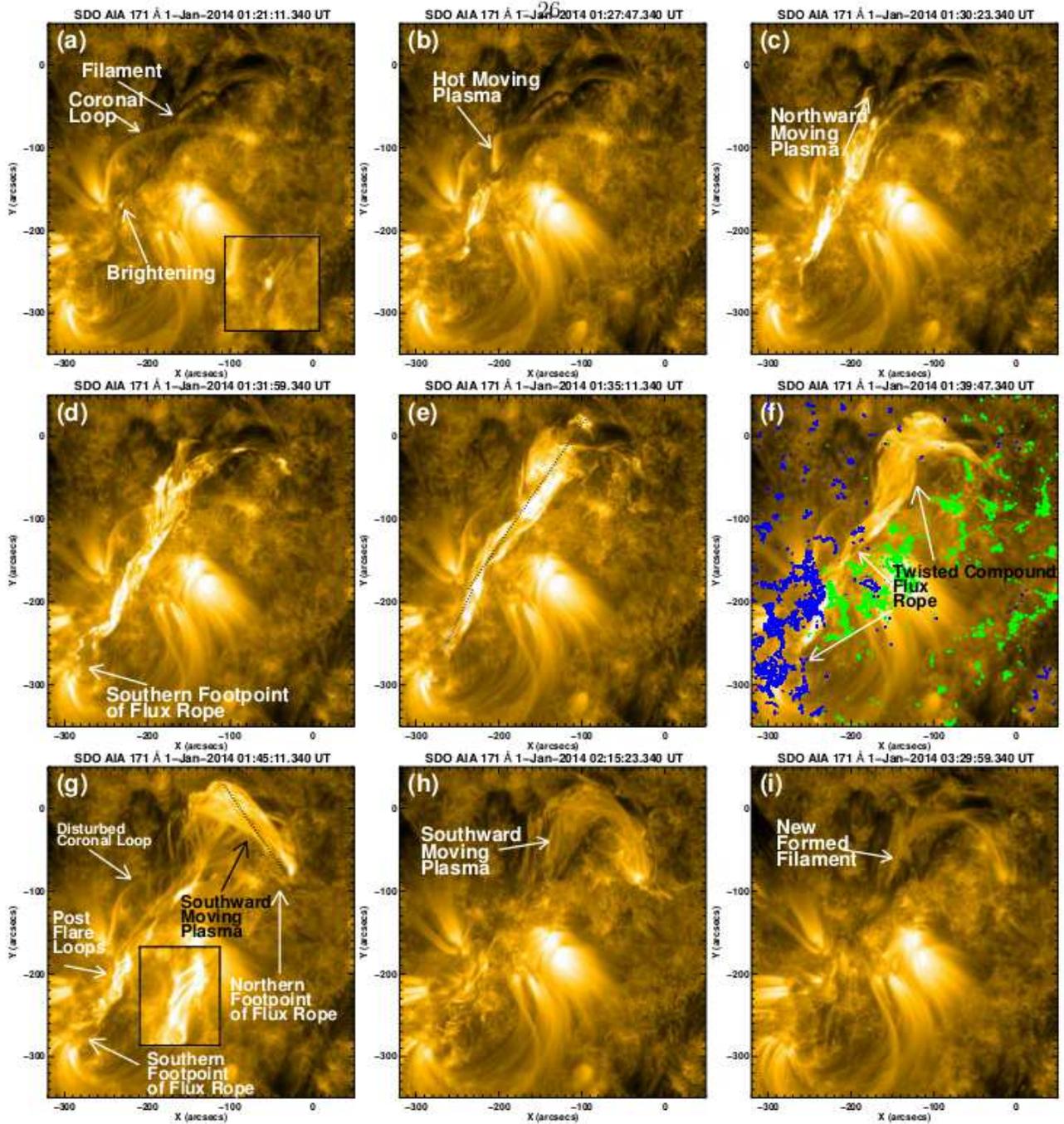}
	}
\vspace*{-4cm}
\caption{Selected SDO/AIA 171 \AA~images showing the overall dynamics of twisted flux rope in coronal temperature. Dotted black/blue lines represent the approximate trajectories along which the distance-time plot of northward and southward moving hot plasma have been measured. The green/blue contours in panel (f) shows the positive/negative polarity regions. The contour levels are $\pm100$, $\pm200$, $\pm400$ Gauss.} 
\label{}
\end{figure}
\clearpage
\begin{figure}
\vspace*{-2cm}
\centerline{
	\hspace*{0.0\textwidth}
	\includegraphics[width=1.2\textwidth,clip=]{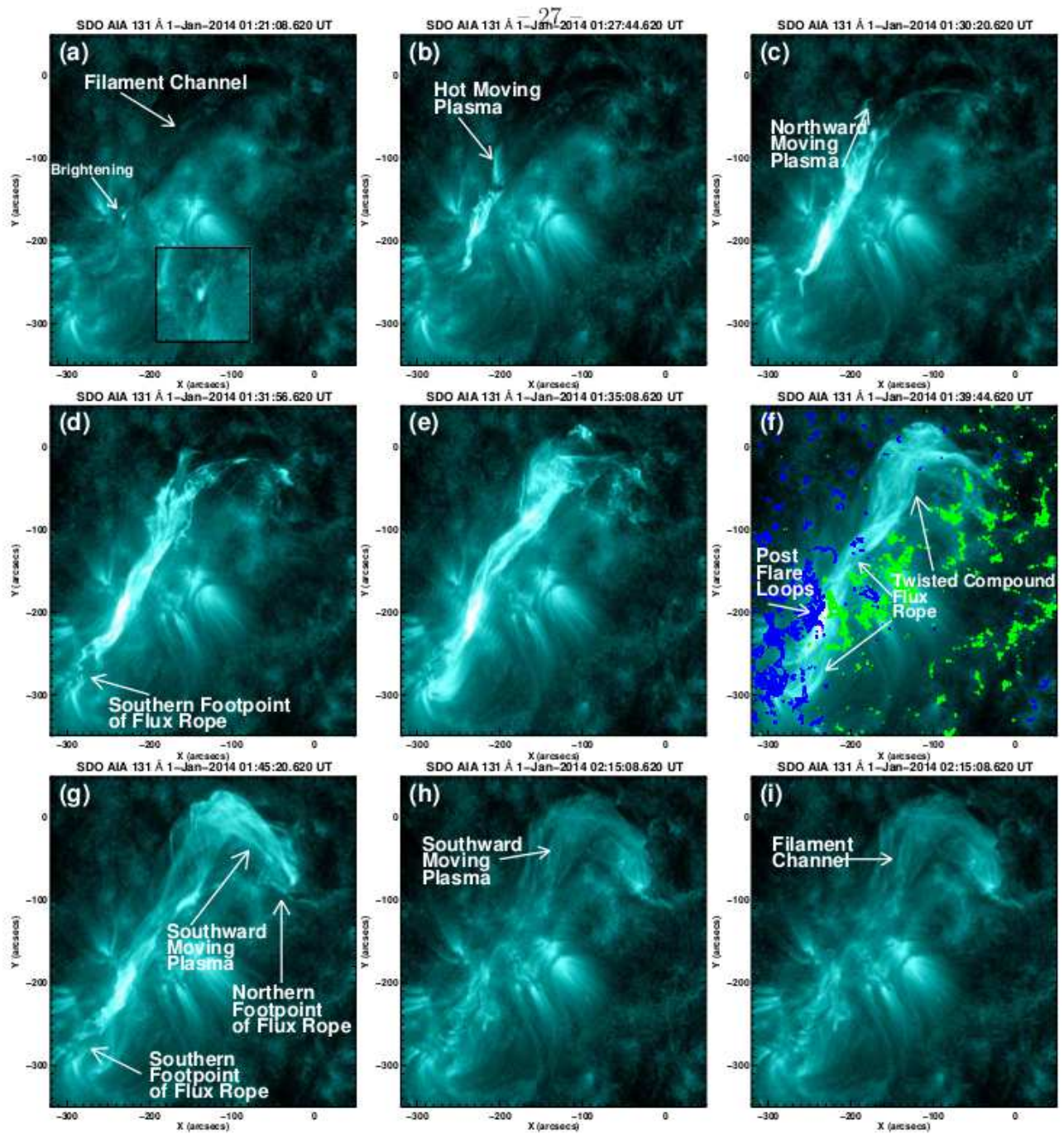}
	}
\vspace*{-4cm}
\caption{Selected SDO/AIA 131 \AA~images showing the overall dynamics of coronal plasma at hot coronal temperature. The green/blue contours in panel (f) shows the positive/negative polarity regions. The contour levels are $\pm100$, $\pm200$, $\pm400$ Gauss.} 
\label{}
\end{figure}
\clearpage
\begin{figure}
\vspace*{-2cm}
\centerline{
	\hspace*{0.0\textwidth}
	\includegraphics[width=1.2\textwidth,clip=]{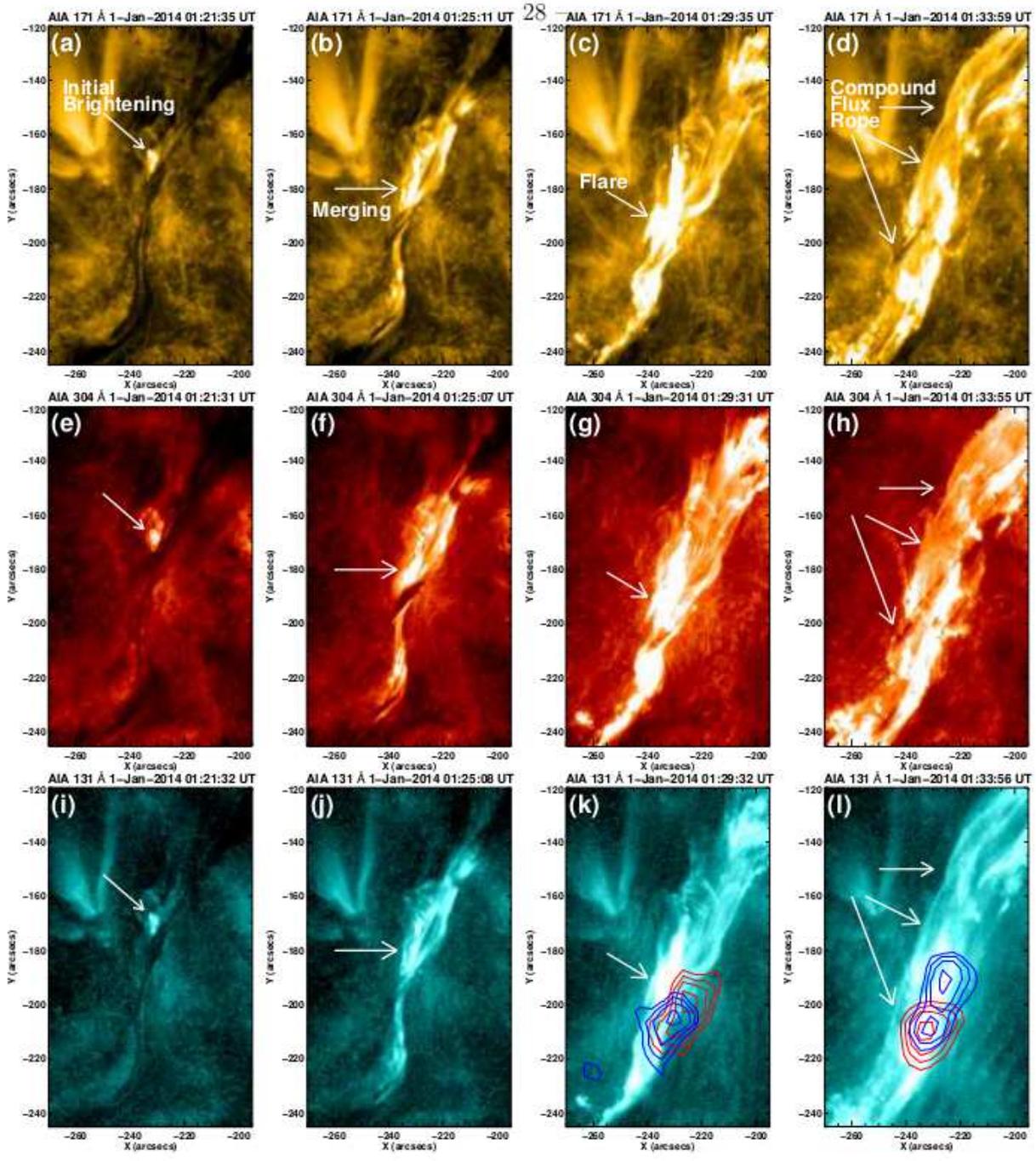}
	}
\vspace*{-4cm}
\caption{Selected SDO/AIA 171, 304 and 131 \AA~images showing the flux ropes merging dynamics. In the last panel the contours are the RHESSI X-ray contours at 6-12 keV (red) and 12-25 (blue) keV energy bands. The contour levels are 60\%,70\%,80\%, 95\% of peak intensity. The integration time is 20 seconds.} 
\label{}
\end{figure}
\clearpage
\begin{figure}
\vspace*{-5cm}
\centerline{
	\hspace*{0.0\textwidth}
	\includegraphics[width=1.2\textwidth,clip=]{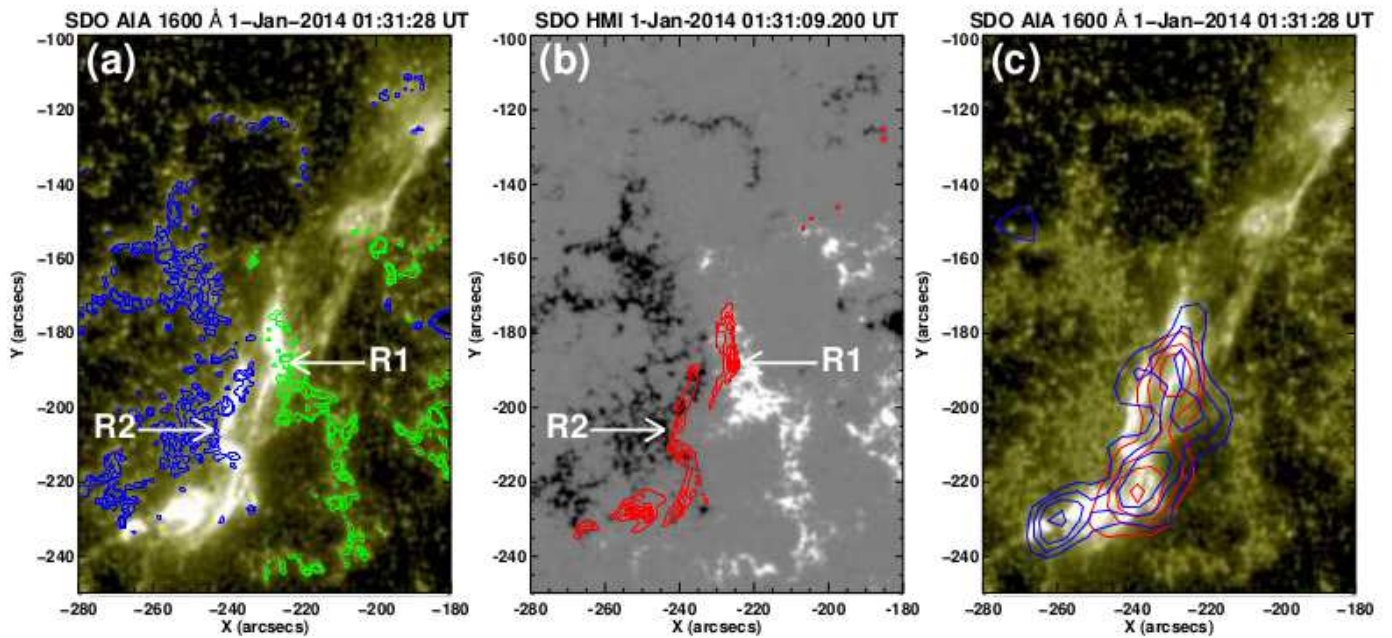}
	}
\vspace*{-9cm}
\caption{(a) SDO/AIA 1600 \AA~image overplotted by the SDO/HMI contours. Green/blue contours represent the positive/negative polarity regions. The contour levels are $\pm200$, $\pm400$ Gauss. (b) SDO/HMI image overplotted by the SDO/AIA 1600 \AA~intensity contours. These red contours show the two ribbons of the flare. The contour levels are 12\%,20\%,30\%,40\%,50\%,60\%,70\%,80\%,90\% of peak intensity. Northern and southern flare ribbons are represented by R1 and R2 respectively. (c) SDO/AIA 1600 \AA~image overplotted by the RHESSI X-ray contours at 6-12 keV (red) and 12-25 (blue) keV energy channels. The contour levels are 60\%,70\%,80\%, 95\% of peak intensity. The integration time is 20 seconds.}
\label{}
\end{figure}
\clearpage
\begin{figure}
\vspace*{-5cm}
\centerline{
	\hspace*{0.0\textwidth}
	\includegraphics[width=1.2\textwidth,clip=]{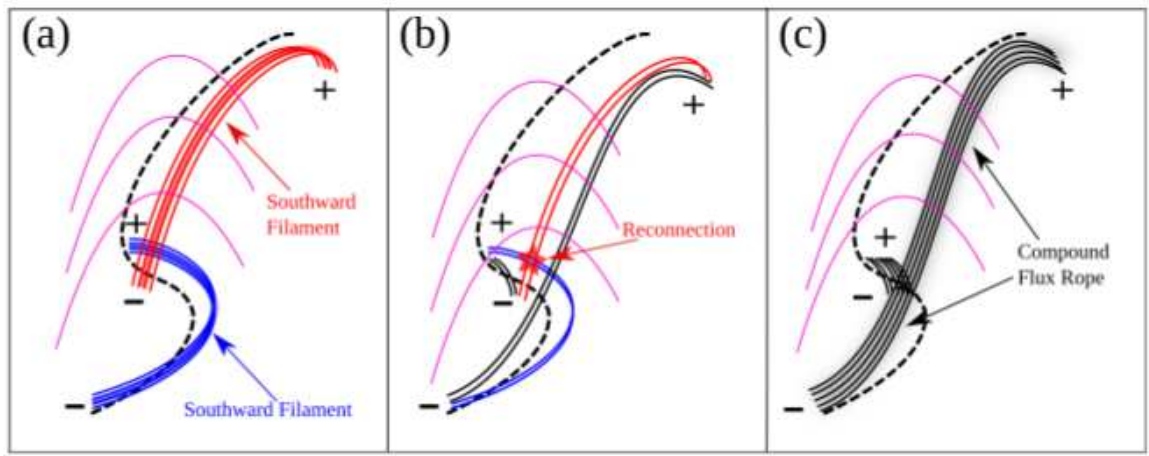}
	}
\vspace*{-9cm}
\caption{Schematic representation showing the interaction of two filament channels field lines and formation of compound flux rope via tether cutting reconnection. Black and blue lines show the field lines before interaction, while the black lines show the field lines after merging/reconnection. Dotted black line shows the polarity inversion line. Pink lines are the representation of overlying fields.}
\label{}
\end{figure}
\clearpage
\begin{figure}
\vspace*{-5cm}
\centerline{
	\hspace*{0.0\textwidth}
	\includegraphics[width=1.2\textwidth,clip=]{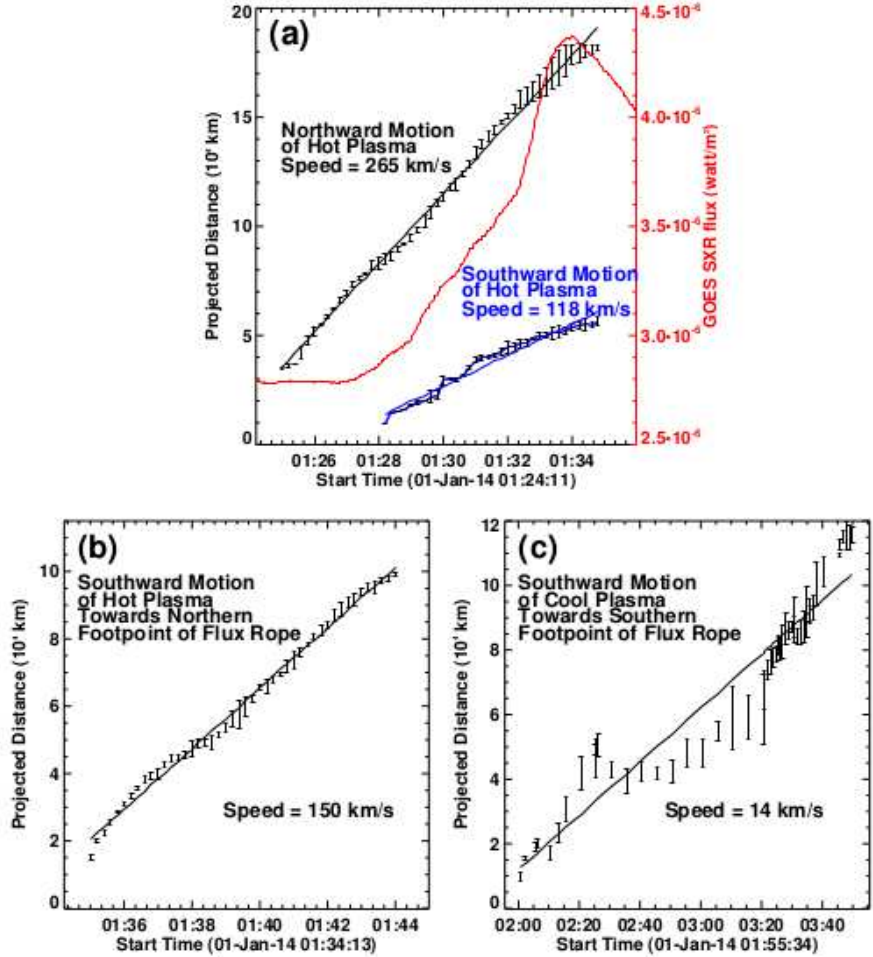}
	}
\vspace*{-6cm}
\caption{Distance-time measurement of the northward (black color curve) and southward (blue color curve) moving plasma (a), moving plasma toward northern foot point of the compound flux rope (b) and the southward moving plasma (c). Red curve in panel (a) shows the GOES X-ray flux in 1-8 \AA~wavelength band. Measurements for the plots (a) and (b) are made using AIA 171 \AA~images, while for the plot (c) is made using GONG $H\alpha$ images. The value of error bars are calculated as a standard deviation of the three repeated measurements. The speeds are calculated using the linear fit to these data points. The approximate trajectories along which the distance time measurement has been made are shown in Figures 5e, 5g and 3i respectively.}
\label{}
\end{figure}
\clearpage
\begin{figure}
\vspace*{-5cm}
\centerline{
	\hspace*{0.0\textwidth}
	\includegraphics[width=1.2\textwidth,clip=]{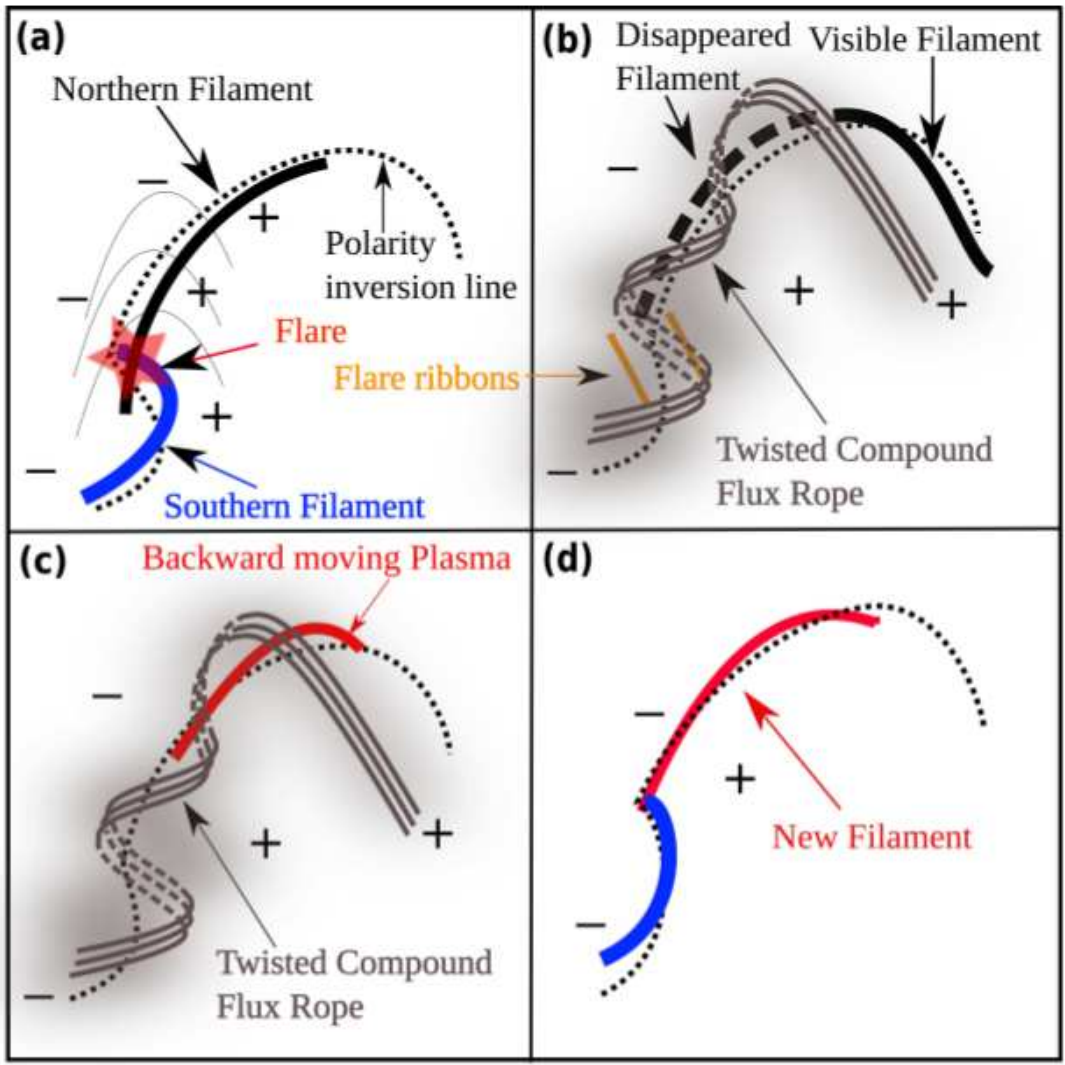}
	}
\vspace*{-5cm}
\caption{Schematic of the filaments cool and hot plasma dynamics within the compound flux rope. Dotted black line represent the polarity inversion line. Thick black and blue lines represent the northward and southward filaments. Red line represents the reformed filament. Twisted flux rope are represented by the brown color, with solid/dashed lines show the front and back side part of field lines around the flux rope. Red patch in panel (a) shows the brightening (i.e., flare) at the crossing. Flare ribbons are represented by the yellow lines below the flux rope in panel (b).}
\label{}
\end{figure}
\clearpage
\begin{figure}
\vspace*{-5cm}
\centerline{
	\hspace*{0.0\textwidth}
	\includegraphics[width=1.2\textwidth,clip=]{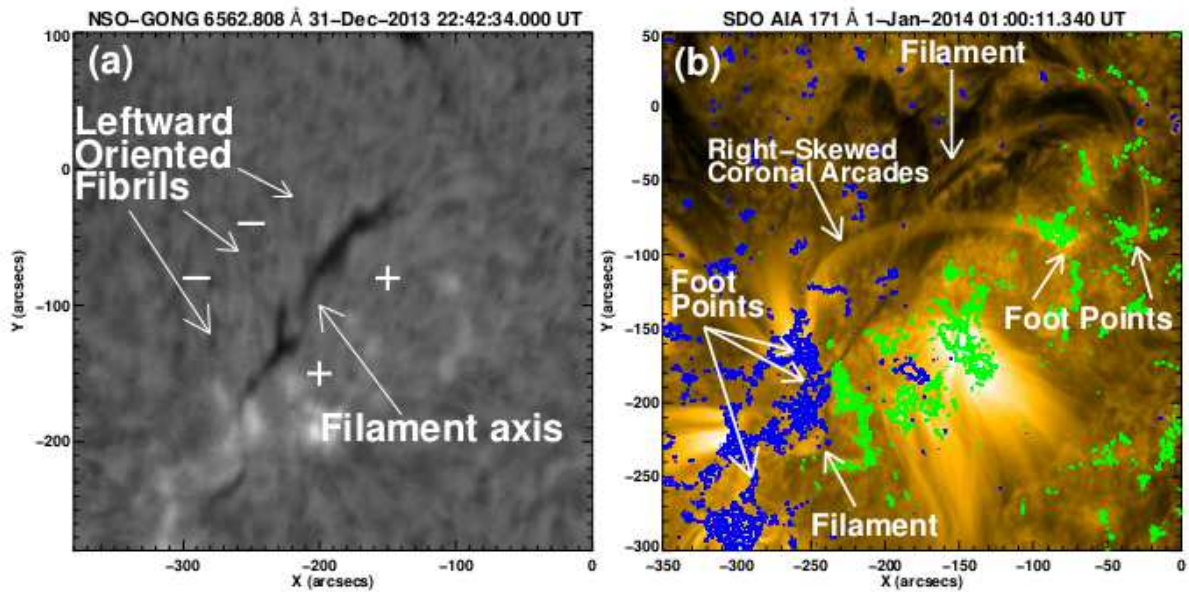}
	}
\vspace*{-9cm}
\caption{(a) GONG $H\alpha$ image at 22:42:34 UT on 2013 December 31 showing the direction of fine chromospheric fibrils emanate towards left when seen from positive polarity side. (b) SDO/AIA 171 \AA~image overlaid by the SDO/HMI contours at 01:00:11 UT. This image showing that the southern footpoint of filaments lie on the negative polarity, while the northern footpoints lie in the positive polarity region. The overlying right-skewed arcade is also shown by white arrow. All these observations provide the evident of sinistral chirality of filaments. The contour levels are $\pm100$, $\pm200$, $\pm400$ Gauss.}
\label{}
\end{figure}
\clearpage
\begin{figure}
\vspace*{-5cm}
\centerline{
	\hspace*{0.0\textwidth}
	\includegraphics[width=1.2\textwidth,clip=]{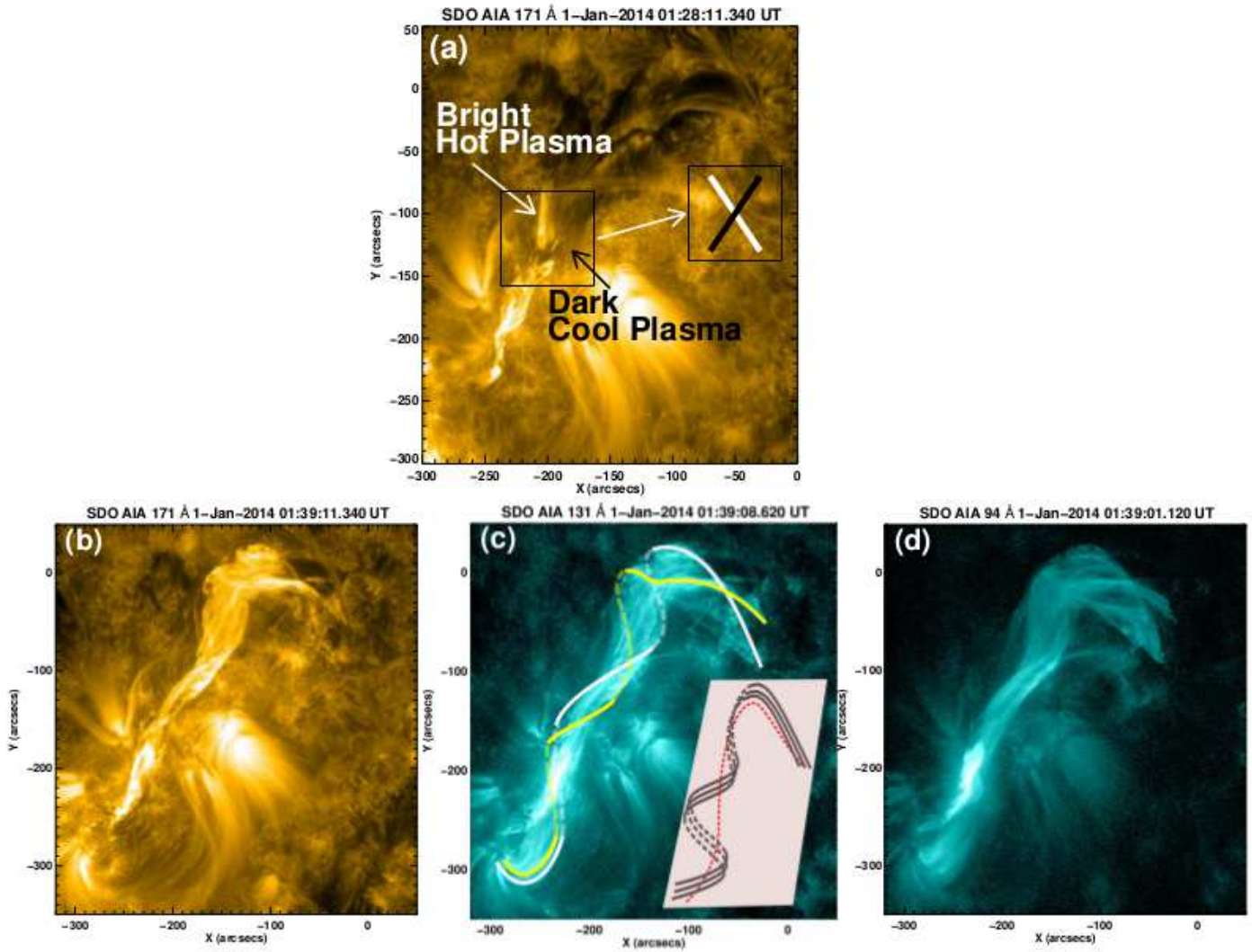}	
	}
\vspace*{-6cm}
\caption{(a) SDO/AIA 171 \AA~image at 01:28:11 UT showing the moving bright hot plasma under the dark cool filament plasma material. The schematic is also overplotted in this figure. (b) SDO/AIA 171, 131 and 94 \AA~images at $\approx$01:39 UT showing the right handed twisted flux rope. Schematic is also presented over the panel (c). The two visible flux rope field lines are represented in the panel with white and yellow lines. Solid/dotted lines showing the part of field lines over and under the flux rope axis. All these figures show the evidence of positive helicity of the surrounding flux rope.}
\label{}
\end{figure}
\clearpage
\begin{figure}
\vspace*{-5cm}
\centerline{
	\hspace*{0.0\textwidth}
	\includegraphics[width=1.2\textwidth,clip=]{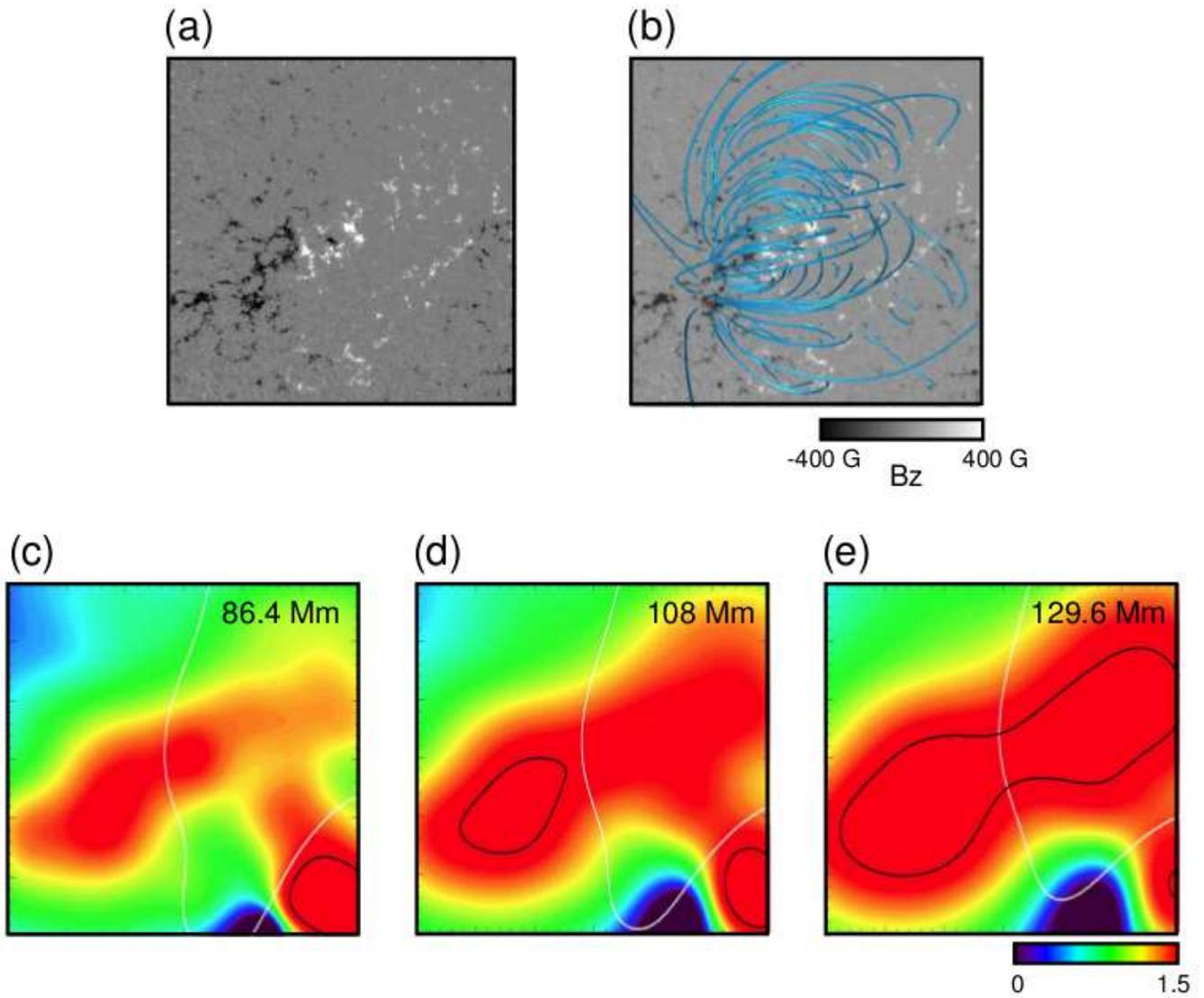}
	}
\vspace*{-6cm}
\caption{Top panel: Radial component of magnetic field around the active region NOAA AR 11938 near the filament axis (a) and the potential field extrapolated over it (b). Bottom panel: Measured decay index at 86.4 Mm (c), 108 Mm (d) and 129 Mm (e) heights. White and black lines indicate the polarity inversion line and contour of decay index 1.5 respectively.}
\label{}
\end{figure}
\clearpage
\begin{figure}
\vspace*{-5cm}
\centerline{
	\hspace*{0.0\textwidth}
	\includegraphics[width=1.2\textwidth,clip=]{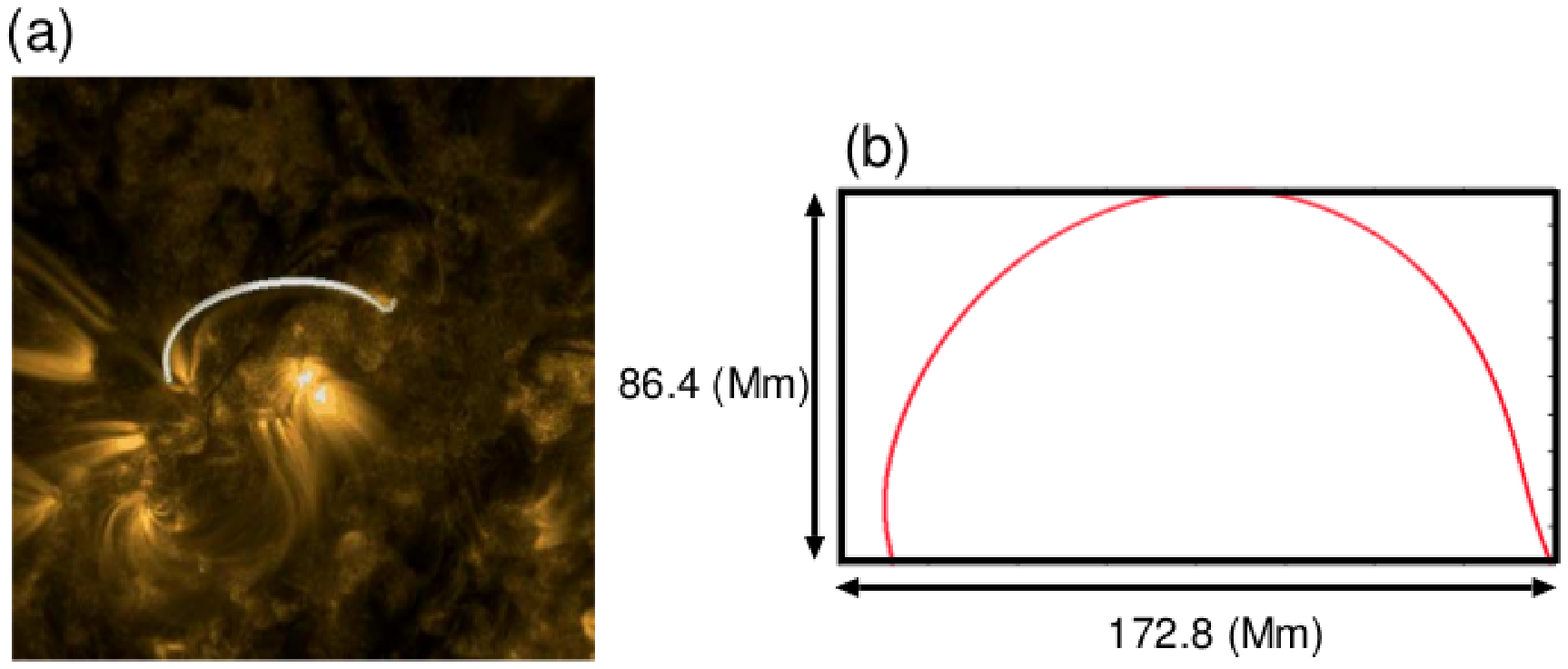}
	}
\vspace*{-9cm}
\caption{SDO/AIA 171\AA~image overplotted by the potential field line matching best with the overlying arcade (a) and the estimation of height of this potential field line (b).}
\label{}
\end{figure}
\end {document}